\def\deg{\mbox{$^{\circ}$}}

\newcommand\JMR{ {J.~Magn.~Reson.} }
\newcommand\JMRA{ {J.~Magn.~Reson.~Series A} }
\newcommand\JMRB{ {J.~Magn.~Reson.~Series B} }
\newcommand\MRM{ {Magn.~Reson.~Med.} }
\newcommand\vol[1]{{#1}}

\newcommand\bm[1]{\mbox{\boldmath $#1$}}
\newcommand\Eq[1]{Eq.$\:$[\,\ref{#1}\,]}
\newcommand\Fig[1]{Fig.$\:$\ref{#1}}

\newcommand\musec{$\mu$s}

\newcommand\T[1]{\mbox{$T_{\rm #1}$}}
\newcommand\RFmax{$RF_{\rm max}$}

\documentclass[3p, twocolumn]{elsarticle}

\usepackage{amssymb}
\usepackage{graphicx}
\usepackage{lineno}

\usepackage{amssymb}
\usepackage{graphicx}
\usepackage{lineno}

\begin{document}

\begin{frontmatter}



\title{Optimal control design of band-selective excitation pulses \\
that accommodate relaxation and RF inhomogeneity}

\author[a]{Thomas E. Skinner\corref{cor1}} \ead{thomas.skinner@wright.edu}\cortext[cor1]{Corresponding author.}
\author[a]{Naum I. Gershenzon}
\author[b]{Manoj Nimbalkar}
\author[b]{Steffen J. Glaser} \ead{glaser@tum.de}

\address[a]{Physics Department, Wright State University, Dayton, OH 45435, USA}
\address[b]{Department of Chemistry, Technische Universit\"at  M\"unchen, Lichtenbergstr. 4, 85747 Garching, Germany}

\begin{abstract}
Existing optimal control protocols for mitigating the effects of relaxation and/or RF inhomogeneity on broadband pulse performance are extended to the more difficult problem of designing robust, refocused, frequency selective excitation pulses.  For the demanding case of \T{1} and \T{2} equal to the pulse length, anticipated signal losses can be significantly reduced while achieving nearly ideal 
frequency selectivity.  Improvements in performance are the result of allowing residual unrefocused magnetization after applying relaxation-compensated selective excitation by optimized pulses (RC-SEBOP).  We demonstrate simple pulse sequence elements for eliminating this unwanted residual signal.

\end{abstract}


\end{frontmatter}


\section{Introduction}
Frequency-selective pulses have widespread utility in magnetic resonance and have motivated significant efforts towards their design \cite{Hill73,Bod76,Freeman78,Hoult79,Bod82,Ernst83,Freeman84,Silver&Hoult84,Freeman87,Hardy88,Warren88,Canet88,Freeman89a,Freeman89b,Bod89,Yagle90,Shinnar89a,Shinnar89b,Shinnar89c,Shinnar89d,Carlson91,Freeman91a,Freeman91b,Freeman92,Carlson92,Rourke92,Bod92,K&F93,Rourke94,Emsley94,K&F95,Kupce95,Emsley95,Boehlen96,Emsley96,Lunati98,Virgili98,Freeman98,Freeman99,Mueller99,Canet99,Canet00,Sterk01,Emsley02,Shaka02,Griffin04}.  In many useful cases, the resulting methodologies can achieve the best approximation to the ideal rectangular profile for spin response as a function of frequency offset.

However, in all of these approaches to pulse design, performance criteria that can be included in the design protocol are restricted either by analytical procedures of highly specific scope or by numerical methods that are limited by the efficiency of the optimizations employed.
As a result, pulse response is typically optimized only for the nominal ideal RF pulse values. 
In addition, although the length of pulses required for narrowband applications can significantly reduce their effectiveness when relaxation times are comparable to the pulse length \cite{Hajduk93, Raddi2000}, the solution to the problem---selective pulses which are less sensitive to relaxation effects---can also be demanding for standard design methods \cite{Nuzillard94, Shen95, Kupce95, Rourke04, Rourke07, Issa09}. 

To make these design challenges tractable, the space of possible pulse shapes is often reduced by forcing the solution to be a member of a particular family of functional forms (for example, finite Fourier series).  Thus, potentially, there are important solutions that are missed.

Over the past decade, we have shown optimal control theory to be an efficient and powerful method that can be applied to a wide range of challenging NMR pulse design problems without restricting the space of possible solutions \cite{Khaneja01,Reiss02,Skinner03,Khaneja03,Skinner04,Kobzar04,Skinner05,Grape,Kobzar05,Skinner06,Neves06,Gershenzon07,Gershenzon08,Kobzar08,Neves09,Braun10,Glaser10,Skinner11,Glaser11}.
It is capable of designing broadband pulses \cite{Gershenzon07} and selective pulses \cite{SkinnerENC07, GershenzonENC07} that are simultaneously tolerant to RF inhomogeneity and relaxation effects, which we develop further in the present work.  

\section{Selective pulse design}

Optimal control (including similar, related optimization procedures) was originally introduced into magnetic resonance for the design of band-selective pulses, primarily for imaging \cite{Lurie85, O'Donnell85, Conolly86, Mao86, Murdoch87, Rosenfeld96, Rosenfeld97}. It was quickly supplanted by the efficient Shinnar-LeRoux (SLR) algorithm \cite{Shinnar89a,Shinnar89b,Shinnar89c,Shinnar89d,Pauly91}, which establishes a correspondence between frequency-selective pulse design and digital filter design.  There are fast, non-iterative algorithms for the ideal filter and, hence, the ideal pulse.  Unfortunately, the algorithm does not accommodate additional criteria, such as tolerance to RF inhomogeneity (included in some of the earliest optimal control-related approaches \cite{Lurie85, Murdoch87}) or relaxation effects.  In addition, the most applicable and widely used form of the algorithm derives pulses which produce a specific linear phase dispersion in the spectral response.  Pulses producing no phase dispersion, suitable for spectroscopy, are more problematic for the SLR algorithm. 

We first provide an overview of well-known issues relevant to selective pulse design, since there is considerably less freedom in the choice of parameters compared to broadband pulses.  For example, in designing broadband pulses, we have shown \cite{Kobzar04} there is at best only a marginal relation between the maximum amplitude, \RFmax, of a pulse and the achievable excitation or inversion bandwidth, as long as the pulse length, \T{p}, is allowed to increase sufficiently.  Alternatively, increasing \RFmax\ for a given \T{p}\ can improve performance for a given bandwidth or increase the bandwidth. There can also be innumerable broadband pulses that provide approximately equal performance for a given $RF_{\rm max}$, $T_p$, and bandwidth.  

Selective pulses are far more constrained, with a well-known relation between the selective bandwidth and \T{p}, and much tighter limits on the choice of \RFmax\ for a given product of bandwidth and \T{p} \cite{Pauly91}.  We provide only a simple overview of the optimal control methodology, emphasizing the modifications necessary for the present work.  The basic algorithm for optimizing pulse performance over a range of resonance offsets and RF inhomogeneity is described fully in  \cite{Skinner03}.  Details related to incorporating relaxation \cite{Gershenzon07} and specific dispersion in the phase of the final magnetization\cite{Skinner05, Gershenzon08, James91a}, which we refer to as the phase slope, are provided in the associated references. 

\subsection{Phase slope}
\label{PhaseSlope}

Values of the phase slope, $R$, at each offset \cite{Gershenzon08} characterize the net phase dispersion that accumulates during a pulse of length $T_p$.  The phase slope is defined relative to the net precession of transverse magnetization that would be produced solely by chemical-shift evolution during the same time interval, $T_p$.  A pulse that produces focused
magnetization of fixed phase for all spins in the offset range of
interest would have constant $R = 0$ (i.e., a self-refocused pulse).
Many selective pulses are symmetric, $R = 1/2$ pulses \cite{Freeman84,Conolly86,Pauly91,Rourke92,K&F93,Rourke94}. The symmetry of the resulting pulse provides
an advantage in the development of various algorithms used
in selective pulse design, such as SLR, inverse
scattering \cite{Carlson91,Rourke92}, polychromatic \cite{K&F93}, and stereographic projection
\cite{Rourke94}. In fact, the standard form of the SLR algorithm \cite{Pauly91} can only
generate linear phase of this value.  Sophisticated algorithms allowing for more general phase in selective pulses are described in the literature \cite{Carlson92,Schulte07,Schulte08}, but they are specific to this particular performance factor and cannot include tolerance to variations in other experimentally important parameters.

By contrast, including additional performance criteria, such as a general phase response, is straightforward for optimal control. Initial magnetization $\bm{M}(t_0)$ is driven by the RF controls to a final magnetization $\bm{F}$ that is defined for each resonance offset in the desired range.  To excite transverse magnetization of linear phase slope $R$, we consider target states for each offset $\omega$ in the excitation bandwidth of the form \cite{Gershenzon08}
	\begin{equation}
\bm{F} = [\,\cos(\varphi), \sin(\varphi), 0 \,]
\label{F(R)}
     \end{equation}
Choosing $\varphi = R\,\omega T_p$ gives a linear phase slope, but any function can be considered to define a useful target phase, such as quadratic or higher order. 

Selective excitation most simply requires changing the target to $\bm{F} = [\, 0, 0, 1\,]$ for offsets outside the desired bandwidth.  In principle, this stopband includes an infinite range of frequencies that must therefore be truncated at some chosen value.  We found as a practical matter that choosing the stopband to be $\sim 5$ times the passband width was sufficient to eliminate excitation at higher frequencies for the pulse parameters used here.  
This value can easily be adjusted upwards if necessary, or, alternatively, high-frequency components of the resulting pulse determined from Fourier analysis can be deleted after verifying they have no significant effect on the passband excitation.

In addition, since the ideal rectangular offset response can not be excited by a finite-length pulse, there must be a transition connecting the excitation frequencies to the nulled frequencies.  This is typically defined in terms of design parameters for finite impulse response (FIR) filters.  An overview of the issues relevant to our optimal control implementation follows.

\subsection{Selective pulses as digital filters}

For design conditions employing ideal RF in the absence of relaxation, selective pulse performance is completely determined by the desired passband width $B$, pulse length $T_p$, transition width $W$ joining the passband and stopband, and residual signal fluctuation or ripple $\delta_1$ and $\delta_2$ about the ideal target amplitude for the passband and stopband, respectively.  

The passband frequency $\nu_p$ and stopband frequency $\nu_s$ are defined where the magnitude of the magnetization response becomes less than the associated fluctuations $1-\delta_1$ and $|\delta_2|$, as illustrated in Fig.1 (adopted from Ref.~\cite{Pauly91}).  The frequency where the amplitude drops to one-half is approximately the average of these two frequencies.  The full width of the filter is defined as twice this value, giving a bandwidth $B = \nu_s + \nu_p$ and a fractional transition width $W = (\nu_s - \nu_p)/B$.

More specifically (and again emphasizing the design conditions stated at the beginning of the section), selective pulse performance is constrained by  relations for optimal FIR filters of the form 
	\begin{equation}
W T_p B = f(\delta_1, \delta_2),
\label{TBW}
     \end{equation}
in terms of an empirically derived function $f(\delta_1, \delta_2)$ \cite{Rabiner}.  For a given value of $f = W T_p B$, smaller (larger) $\delta_1$ gives larger (smaller) $\delta_2$.  Alternatively, for fixed $\delta_1$ or fixed $\delta_2$, values of $f$ increase as $\delta_2$ or $\delta_1$, respectively, decrease.  Flexibility in selective pulse design is thus purchased at the cost of trade-offs among bandwidth, pulse length, transition width, and ripple amplitudes.  Choosing any four of the set determines the fifth.  

This relation appears to have been little used in the spectroscopic community.  
Although the precise form of the function $f(\delta_1, \delta_2)$ holds only for $R=1/2$ pulses, we have found it to be a useful qualitative indicator for more general $R$.  One important implication is that pulse performance for a given absolute transition width $BW = \nu_s - \nu_p$ can be made independent of the passband width, $B$. Fixed \T{p} results in the same performance in terms of residual signal (ripple) for different $B$ as long as the transition width $B W$ is constant.  This was observed empirically and noted in \cite{Griffin04}.  We thus use \Eq{TBW} to inform our optimal control design. 

\subsection{Optimal control}

The approximation to the ideal rectangular frequency response profile, as illustrated in Fig.1, can be readily obtained using a variety of methods---among them, the Parks-McClellan (PM) algorithm for linear-phase FIR digital filters \cite{Parks}. After calculating the PM polynomial, the standard SLR algorithm effectively inverts the frequency response to produce a linear phase ($R = 1/2$) RF pulse. 
The equivalent optimal control approach would be to use the polynomial response function as the target response and efficiently modify the RF controls to achieve the allowed performance.  

However, as noted already, the target response derived using the PM algorithm applies only to $R=1/2$, using the ideal RF amplitudes, in the absence of relaxation.  More significantly, optimal control does not need to know the polynomial response.  One can simply define the ideal passband/stopband frequency response, and the algorithm will find the response allowed by the particular choice of bandwidth, pulse length, and transition width.  Different functional forms can be defined for the response in the transition region to provide additional flexibility.  The response at each frequency can also be given weights to fine-tune the final excitation profile.  

We now proceed with the design of more robust selective excitation pulses.  
In what follows, we let the frequency response to the pulse (ie, the target function) be undefined in the transition region.  The pulse has the flexibility to do anything it wants there.  An important addition to the algorithm is an adjustable weight function which changes the weight for each particular offset depending on the deviation of intermediate results from the desired performance.  During a given iteration, if the deviation of magnetization from the target for a particular offset is larger than the allowed ripple amplitude, then the weight for this offset increases; otherwise it decreases.  This is an extension and generalization of the method introduced in \cite{Skinner05}.

\section{Results and Discussion}

We present several examples illustrating the advantages of relaxation-compensated selective excitation by optimized pulses (RC-SEBOP).  As discussed, possible performance improvements include increased tolerance to RF inhomogeneity and compensation for relaxation effects. 

\subsection{Compensation for relaxation and RF inhomogeneity}

The algorithm for generating relaxation compensated broadband pulses (RC-BEBOP) has already been developed \cite{Gershenzon07}.  It only needs inclusion of the modified target functions described in Section \ref{PhaseSlope} for application to selective pulses (RC-SEBOP).  The primary result of that work is that substantial signal gains are possible relative to expected losses from short \T{1}, \T{2} by finding trajectories for the transformed magnetization that minimize these losses, even if this requires a longer pulse length. 
Relaxation losses are minimized by keeping spins near the z-axis and orienting them so that all offsets can be transformed to the transverse plane in a very short time at the end of the pulse.
This solution not only mitigates the effects of transverse relaxation, but short \T{1} becomes an advantage due to repolarization of $z$-magnetization during the relatively long time the spins are aligned close to the longitudinal axis.  The net affect is that almost all the RF power is applied at the end of our relaxation compensated pulses \cite{Gershenzon07}.  This is fortunately also consistent with a particular family of solutions for broadband $R=0$ (refocused) pulses \cite{Gershenzon08}.  

There are also solutions for refocused selective excitation pulses \cite{Griffin04} that employ significant RF power throughout the pulse and therefore do not lend themselves to relaxation compensation.  On the other hand, these pulses do an excellent job of minimizing the residual off-axis magnetization in both the passband and stopband.  Consistent with this result, we find empirically  that minimizing relaxation losses for selective pulses competes with minimizing residual off-axis magnetization.  The trajectories that reduce relaxation effects are not compatible with those that achieve good refocusing.  We therefore consider a strategy that maximizes $x$-magnetization in the passband, minimizes it in the stopband, i.e., the standard procedure, but removes any explicit restrictions on residual $y$-magnetization.  This allows the optimal control algorithm to emphasize relaxation compensation and find solutions with considerably enhanced performance compared to those that give highest priority to minimizing the $y$-component.  As we will show, this undesirable residual magnetization can be eliminated after the pulse without significantly affecting performance.

\subsubsection{Excitation profile effects}

In addition to causing signal losses, relaxation can also degrade the uniformity of the excitation profile \cite{Hajduk93}.  SLURP pulses \cite{Nuzillard94} were developed specifically to obtain more uniform response over the selected bandwidth while accepting attenuation due to short \T{1}, \T{2}.  SLURP-1 pulses address the particular case \T{1} = \T{2}, and were derived for various values of the ratio \T{2}/\T{p}.

For the demanding case \T{p} = \T{1} = \T{2} = 1 ms, theoretical performance of SLURP-1 is compared to RC-SEBOP in \Fig{SLURP} for RF miscalibrations of $-10\%$, 0\%, and +10\%, displayed left to right.  For the ideal RF shown in the middle panel, RC-SEBOP provides a signal gain of 60\%, uniformly preserving greater than 90\%\ of the $x$-magnetization over the optimized bandwidth of 4~kHz, in spite of short $T_1, T_2$ equal to the pulse length.  It is relatively insensitive to RF inhomogeneity of $\pm10\%$.  In all cases, there is improved signal suppression from $M_x$ magnetization in the stopband, and a narrower transition width.  

Analogous investigations have optimized relaxation-compensated pulses by applying simulated annealing to RF waveforms represented using finite Fourier series ($\sim8$ frequency components or less) and no compensation for RF inhomogeneity \cite{Shen95, Issa09}.  They achieve signals of $< 80\%$ for the case $T_p = T_1 = T_2$ over a less selective bandwidth than obtained here.  Some of our performance gains may be due to a more efficient optimization procedure that does not restrict the solution space. However, we expect the largest gains are due to the flexibility of allowing residual $y$-magnetization, $M_y$, which can be quite large, as shown in the bottom panel of \Fig{SLURP}.   This is easily eliminated, as shown later, and allows a much more ideal, rectangular excitation profile than previously considered possible for short \T{1}, \T{2} \cite{Hajduk93},  with minimal loss of signal.  
Alternatively, if we choose to minimize the residual $M_y$ during the optimization, this requires trajectories that sacrifice signal enhancement and selectivity of the passband, consistent with the other cited studies.  
\subsubsection{Excitation using small time-bandwidth product pulses}

A simple strategy for reducing relaxation effects is to reduce the pulse length, but according to \Eq{TBW}, achieving acceptable performance for a narrow excitation bandwidth then becomes more problematic.  A given time-bandwidth product requires trade-offs in the sharpness of the excitation profile (transition width) and the signal variation (ripple) in both the passband and the stopband.  Still, one can optimize performance for a desired low value of the time-bandwidth product, as accomplished in the SNOB family of pulses \cite{Kupce95}.  Including relaxation compensation beyond what is accomplished by a short pulse length alone and including tolerance to RF inhomogeneity provide additional opportunities for improved performance.  

For e-SNOB, \T{p} = 1 ms, selective bandwidth $\pm 1.5$~kHz, we designed a RC-SEBOP pulse incorporating relaxation times \T{1} = \T{2} = \T{p} and tolerance to RF inhomogeneity of $\pm 10\%$.  Pulse shapes are shown in \Fig{eSNOB} along with theoretical performance.  Signal gains of a factor of 2 are obtained with the relaxation-compensated pulse, with a sharper excitation profile and improved suppression of $M_x$ magnetization in the stopband. To minimize relaxation losses, RC-SEBOP delivers most of its RF power at the end of the pulse.  As in the previous example, these enhanced performance features are purchased at the cost of a larger residual $M_y$ compared to e-SNOB.  In the next section, we present methods for selecting only the desired $M_x$ magnetization while maintaining the performance advantages of RC-SEBOP.

\section{Experimental}

All the selective pulses considered so far, both traditional pulses and optimal control pulses, produce significant residual $M_y$ magnetization in the passband at non-ideal RF calibration.  Methods for removing this unwanted signal therefore have more general applicability.

To destroy undesired  $M_y$  after selective excitation in a single-acquistion experiment, a hard $90_{-y}^\circ$ flip-back pulse can be applied to store $M_x$ along the $z$-axis.  A gradient pulse is then employed to dephase the $M_y$ component, followed by a hard $90_y^\circ$ pulse for acquisition of the signal due to $M_x$.  This ``crusher'' sequence, implemented in \Fig{Crusher}a as a more general phase-cycled sequence (to be discussed below), was first tested for the case of unlimited-bandwidth hard pulses by applying the hard pulses on resonance.  The transmitter frequency was shifted only for the RC-SEBOP pulse to measure its off-resonance performance.  This sequence is insensitive to the actual $T_2$ of the sample, since there is minimal relaxation during the short hard pulses and no $T_2$ relaxation of magnetization stored along the $z$-axis during the gradient pulse.  In addition, for typical samples with $T_1 \gg T_2$, there are minimal $T_1$ effects as a result of the sequence.  However, for very short $T_1$, repolarization of stopband magnetization during the gradient pulse can lead to slightly more $M_x$ magnetization in the stopband than expected from the theoretical selectivity profile of RC-BEBOP, previously illustrated in \Fig{eSNOB}.  

To eliminate extra stopband signal in cases where $T_1$ is too short for ideal performance of the original crusher sequence, it can be phase-cycled as shown in the figure.  The two acquisitions add constructively in the passband.  Repolarization at stopband frequencies leads to $+z$-magnetization after the first hard pulse in each cycle which is canceled by addition of the two acquisitions after the second hard pulse.  As a general strategy, this sequence works very well for the case of ideal hard pulses with no bandwidth limitations.  It therefore also works well over a bandwidth of $\sim 8$~kHz for a 25~kHz hard pulse, where the performance of the hard pulse is sufficiently ideal, and may be useful for applications that require only a relatively narrow range of stopband frequencies compared to the available RF amplitude of the hard pulse.

More realistically, all pulses have to be applied at the same transmitter frequency to measure the performance of the sequence at larger resonance offsets.  The effective field of the hard pulses will then have a significant $z$-component, and the $y$-magnetization will no longer remain untouched.  However, the gradient still dephases the transverse magnetization remaining after the first hard pulse.  To a first approximation, there is only $z$-magnetization prior to the second hard pulse, and the phase cycle produces transverse components of opposite sign that cancel on addition.

Experimental off-resonance performance of RC-SEBOP and the phase-cycled crusher sequence is demonstrated in \Fig{Expt1} for the residual HDO signal in a sample of 99.96\% D$_2$O, doped with CuSO$_4$ to relaxation times of $T_1 = 1.345$~ms and $T_2 = 1.024$~ms at 298\deg\ K.  

Signals representing offsets between $-15$~kHz to 15~kHz were obtained in steps of 200~Hz by offsetting the transmitter in this fashion for all applied pulses.
The sequence was first implemented using 10\musec\ hard pulses, which resulted in stopband signal of $\sim5\%$ relative to the centerband.  It was then fine-tuned by increasing the hard-pulse length to 10.68\musec\ in the first cycle.  The resulting selectivity profile is in very good agreement with the simulations for eSNOB and RC-BEBOP performance in \Fig{eSNOB}.  We obtain a signal of 0.83 on resonance using the nominal ideal RF values for RC-SEBOP compared to the theoretical value 0.89.  All values are relative to an ideal signal of 1 for the case of no relaxation.  The small disagreement between experiment and simulation can be attributed to RF inhomogeneity/miscalibration in the flip-back pulse, which will leave some small fraction of the desired passband signal in the transverse plane to be destroyed by the gradient.

Since relaxation compensation functions by keeping spins close to the $z$-axis as long as possible during the pulse, the performance of RC-SEBOP is not highly specific to particular \T{1}, \T{2} values.  Although optimized for \T{1} = \T{2} = 1~ms, the $T_p = 1$~ms RC-SEBOP pulse of \Fig{eSNOB} performs well for much shorter relaxation of $T_1 = 708$~\musec\ and $T_2 = 527$~\musec, as shown in \Fig{Expt2}. A signal of 0.73 is obtained on resonance using the nominal ideal RF values for the pulse and the phase-cycled crusher sequence compared to a theoretical value of 0.81 for on-resonance excitation by RC-SEBOP alone.  The pulse could also be optimized for faster relaxation to improve performance further. 

If a single-acquisition sequence is preferred, the sequence of \Fig{Crusher}b can be used to more completely eliminate stopband signal for the case of short $T_1$.  The first gradient is followed by a hard $180^\circ_y$ pulse.  $M_z$ is flipped to $-z$ where any additional magnetization repolarized during the first gradient pulse can relax to zero during a subsequent delay of the same length as the first.  This extra delay can also be used for additional dephasing of unwanted transverse magnetization by a gradient pulse of opposite sign to the first gradient. The sequence ends with a hard 90\deg$_y$ pulse followed by signal acquisition. However, performance depends more sensitively on any RF inhomogeneity or miscalibration of the hard pulses, since there are now two opportunities to destroy leftover transverse magnetization due to imperfect rotation and storage along the $z$-axis. Off-resonance effects of the hard pulses, which now include a 180\deg\ pulse, further degrade performance.  
For the experiment described in \Fig{Expt2}, a signal of 0.68 is obtained using the single-acquisition sequence of \Fig{Crusher}b.
The scheme is included as an option for narrow passband applications.  In addition, one might want to explore possibilities using broadband, shaped 90\deg\ and 180\deg\ pulses.  These could also be incorporated into a simpler sequence that eliminates the gradient in \Fig{Crusher}a and cycles according to [$\pm y$, $y$, Acq($x, -x$)].  These topics are beyond the scope of the present article.

\section{Conclusion}

\begin{figure*}[h]
\includegraphics[scale=.9]{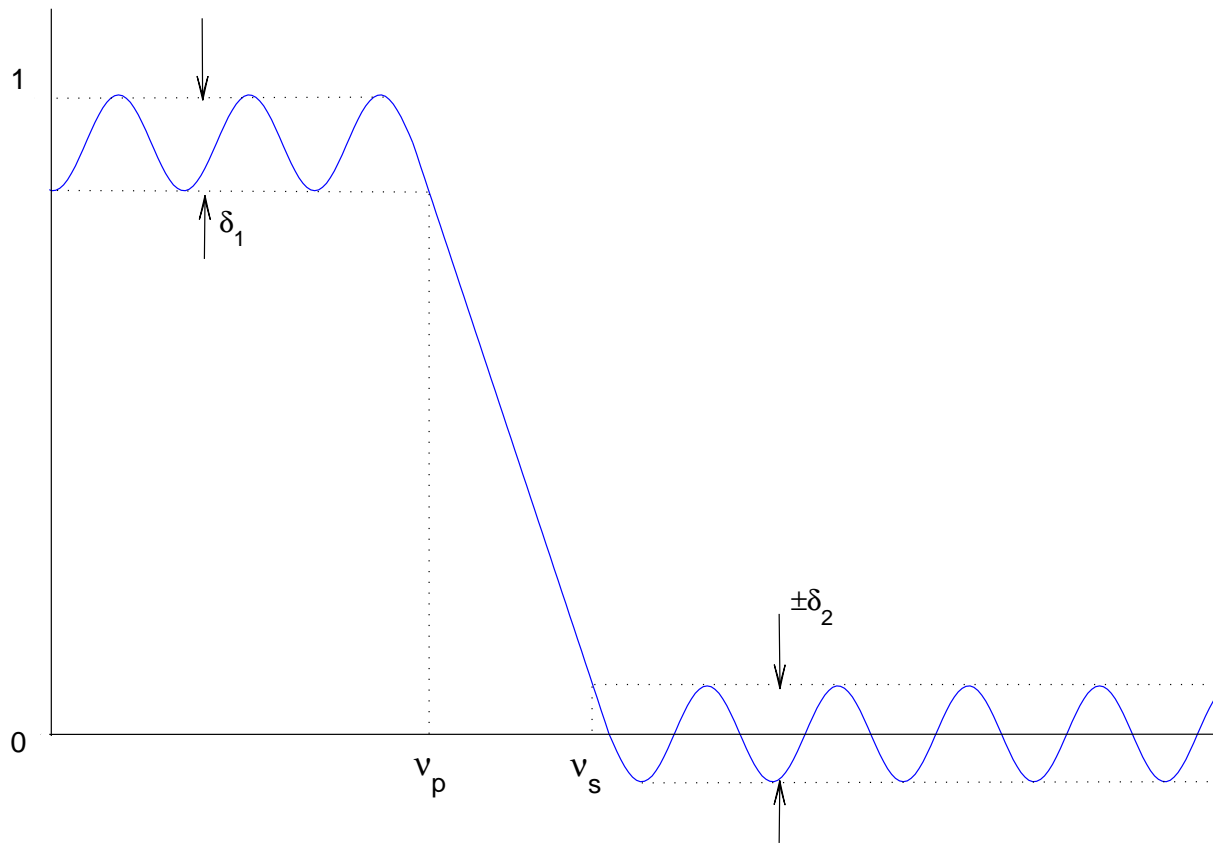}
\caption{(Adapted from Ref.\cite{Pauly91}) A finite length selective pulse can only approximate the ideal rectangular frequency response.  Residual signal or ripple amplitude in the selected frequency spectrum (passband) is denoted by $\delta_1$, with $\delta_2$ representing the ripple over the frequency range where the signal should be nulled (stopband).  The positive frequencies $\nu_p$ and $\nu_s$ define the passband and stopband, respectively. The plotted response is symmetric about the zero frequency.}
\label{ripple}
\end{figure*}

\begin{figure*}[h]
\includegraphics[scale=.8]{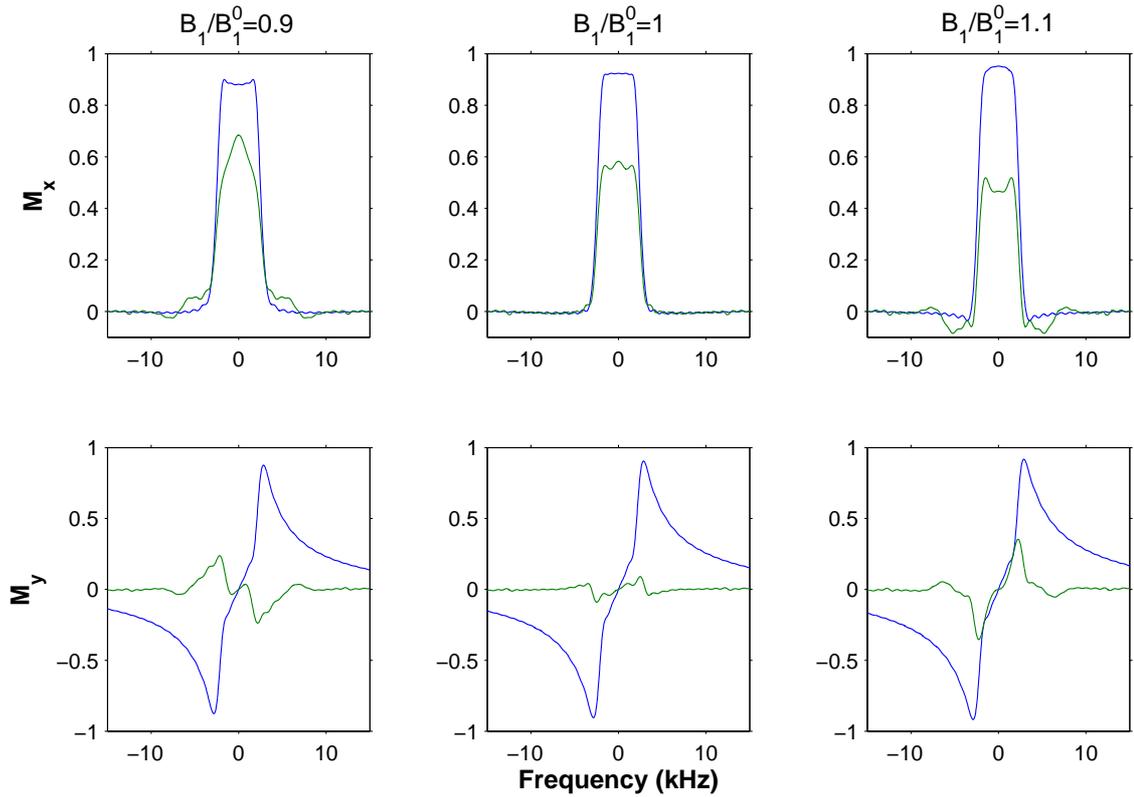}  
\caption{Excited magnetization $M_x$ (top panels) and $M_y$ (bottom panels) is plotted as a function of resonance offset for SLURP-1 (green) and RC-SEBOP (blue) for $T_p = T_1 = T_2 = 1$~ms and
RF inhomogeneity/miscalibration of $-10\%$, 0\%, +10\% reading left to right. Despite these fast relaxation rates, RC-SEBOP preserves $>90\%$ of the desired $M_x$ while achieving a nearly rectangular  profile that is relatively insensitive to $\pm10\%$ RF inhomogeneity over the optimized excitation bandwidth of 4~kHz. The signal gain on resonance for the nominal ideal RF is a factor of 1.6.  Minimal relaxation losses are achieved by allowing large residual $M_y$, particularly in the stopband. This unwanted signal can be subsequently eliminated without significantly affecting performance, as described in the text and Figs.\ref{Crusher}--\ref{Expt2}.}
\label{SLURP}
\end{figure*}

\begin{figure*}[h]
\includegraphics[scale=1]{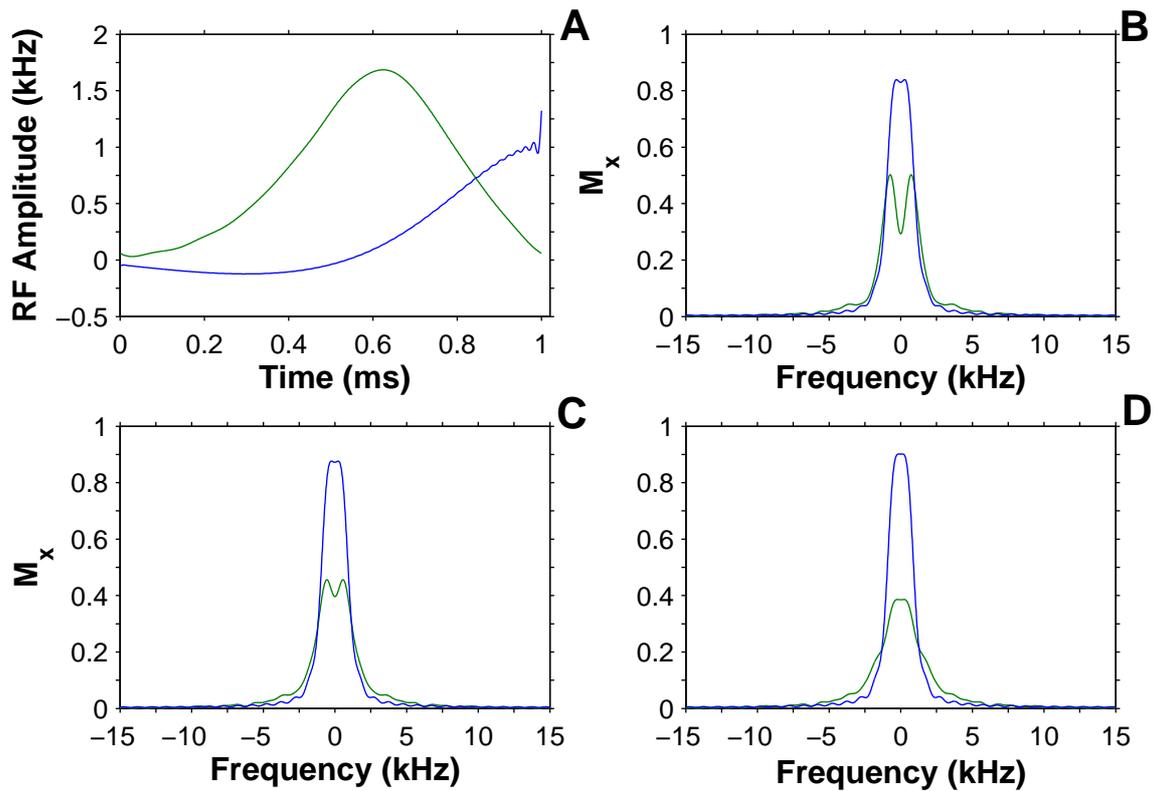}
\caption{Selective pulses eSNOB (green) and RC-SEBOP (blue) of length $T_p = 1$~ms are plotted in panel A.  RC-SEBOP utilizes less RF power, which is applied primarily at the end of the pulse to reduce relaxation losses, as described in the text.  Panels B, C, and D show the frequency response of the pulses for RF miscalibaration/inhomogeneity of $-10\%$, 0\%, and $+10\%$, respectively, for \T{1} = \T{2} = 1~ms.  RC-SEBOP significantly reduces relaxation losses and provides a sharper and more rectangular selectivity profile over the designed excitation bandwidth of 3~kHz.  The signal gain in this example is a factor of 2.}
\label{eSNOB}
\end{figure*}

\begin{figure*}[h]
\includegraphics[scale=2]{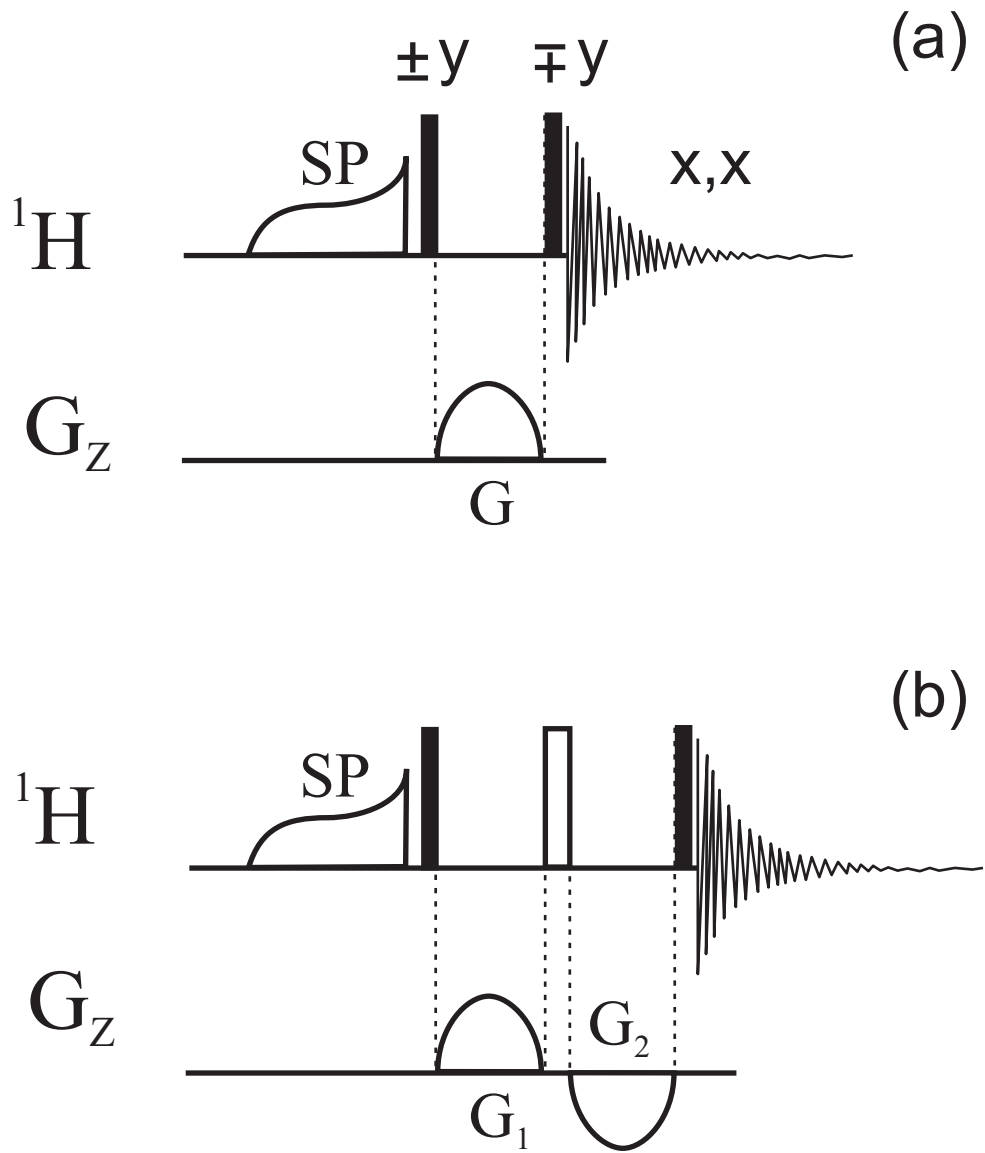}
\caption{(a) Phase-cycled sequence consisting of excitation by selective pulse (SP), hard 90\deg\ pulse, gradient pulse, hard 90\deg\ pulse, and acquisition.  The sequence is designed to eliminate residual $y$-magnetization produced either by standard selective pulses in the presence of RF inhomogeneity or produced by design in the case of RC-SEBOP.  If $T_1$ is sufficiently long so there is no repolarization during the 90\deg\---G---90\deg sequence, the first cycle is not needed, and the second cycle can be employed as a single-acquisition sequence. (b) a single-acquisition sequence designed as an alternative to (a) for the case of short $T_1$.  Further details for both sequences are provided in the Experimental secion.}
\label{Crusher}
\end{figure*}

\begin{figure*}[h]
\includegraphics[scale=.6]{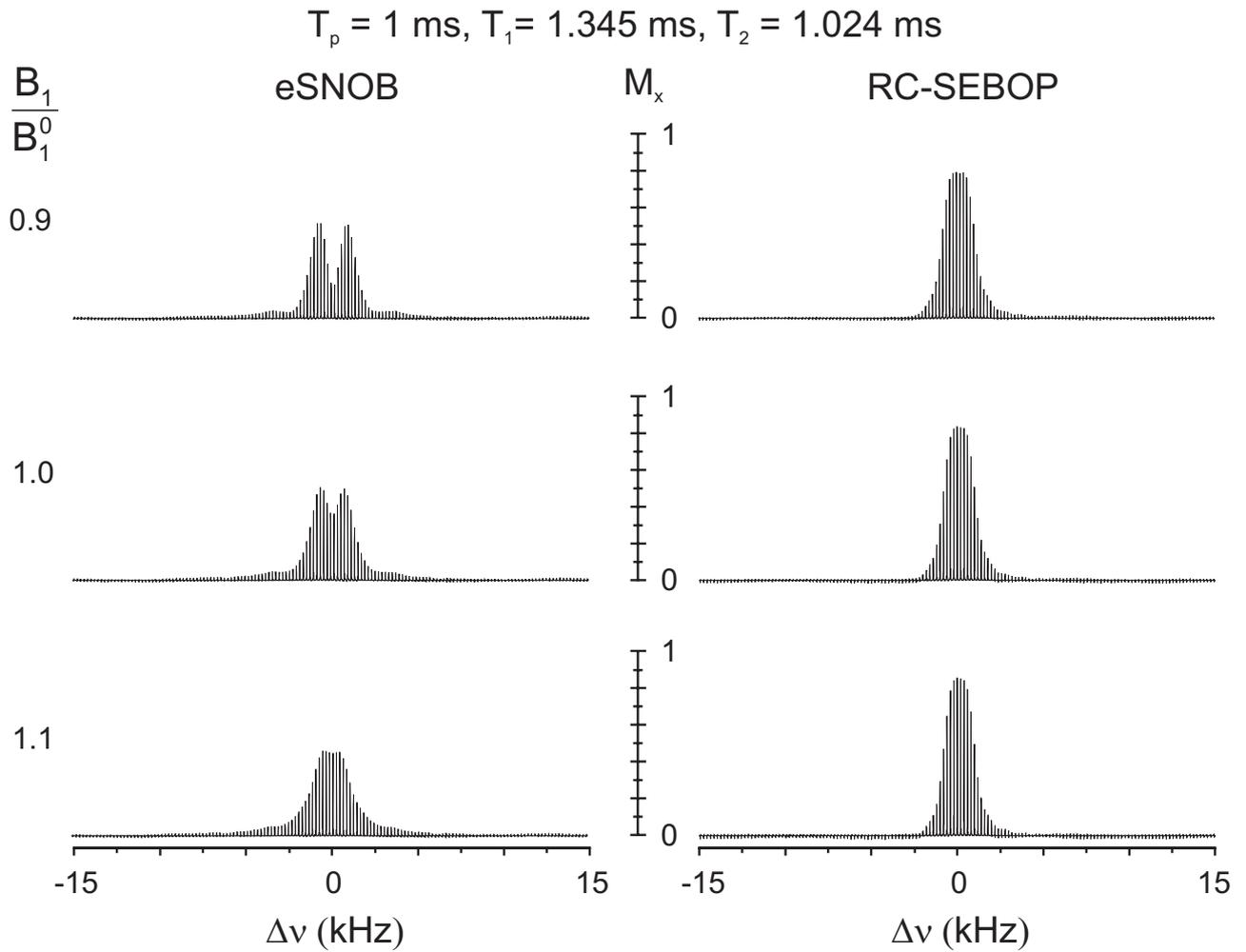} 
\caption{Experimental selectivity profiles of RC-BEBOP (right) and eSNOB (left), pulse length $T_p = 1$~ms, using the phase-cycled crusher sequence of \Fig{Crusher}a applied to a strongly relaxing sample with $T_1$ = 1.345~ms and $T_2$ = 1.024~ms.  Further experimental details are in the text. Results are shown for three values of the RF calibration relative to the ideal value $B_1^0$ for each pulse, showing insensitivity of RC-BEBOP to RF miscalibration of $\pm 10\%$.   Passband signal gains of a factor of 2 are obtained with RC-BEBOP compared to eSNOB.  The combined sequence for relaxation compensation and elimination of $M_y$ preserves 83\%\ of the initial magnetization on-resonance for the case $B_1/B_1^0 = 1$ compared to the theoretical value of 89\%\ for $M_x$ alone shown in \Fig{eSNOB}.}
\label{Expt1}
\end{figure*}

\begin{figure*}[h]
\includegraphics[scale=.6]{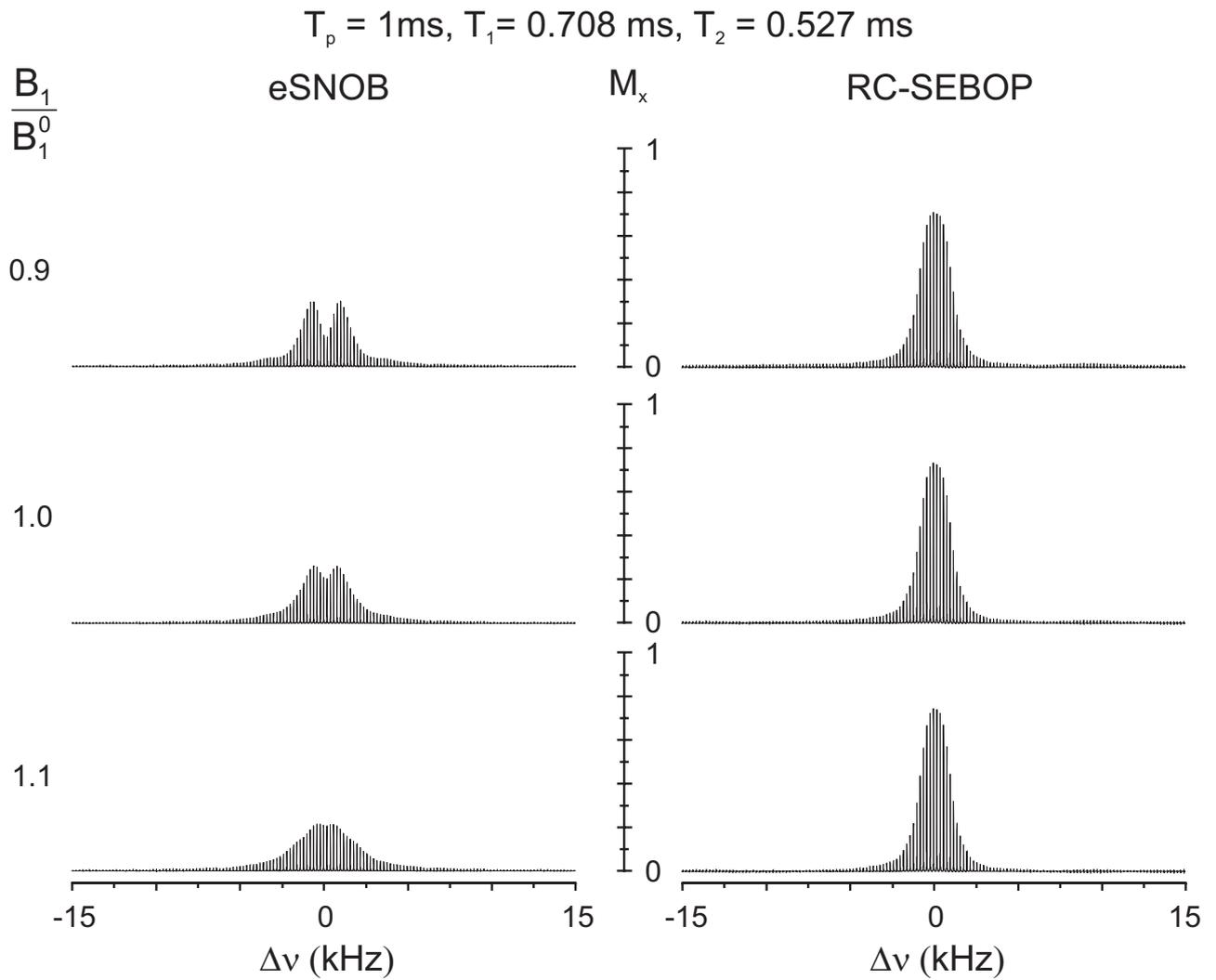} 
\caption{Same as \Fig{Expt1}, but applied to a faster-relaxing sample with $T_1$ = 708~\musec, $T_2$ = 527\musec.  Although RC-SEBOP was optimized for $T_1 = T_2 = T_p = 1$~ms, it provides excellent resistance to relaxation for much shorter values, preserving 75\%\ of the initial magnetization on-resonance for the case $B_1/B_1^0 = 1$.}
\label{Expt2}
\end{figure*}

Pulses which provide robust and enhanced performance despite RF inhomogeneity/miscalibration and relaxation effects are highly desirable.  The optimal control approach to designing refocused selective excitation pulses with these compensatory mechanisms has been presented.  The examples considered the standard selectivity profile comprised of a passband, transition region, stopband, and variations (ripple) in signal uniformity, as illustrated in \Fig{ripple}.  Constraints and trade-offs in the performance among these fundamental parameters were emphasized.  In particular, we found relaxation compensation and null excitation of magnetization in the stopband to be competing goals. Considerable improvements in selectivity and relaxation-compensation for short \T{1}, \T{2} were obtained by allowing residual unrefocused magnetization in both the passband and stopband. This residual magnetization can be readily eliminated without significantly diminishing performance using additional pulse elements.
  

\section*{Acknowledgments}
T.E.S. acknowledges support from the National Science Foundation under Grant CHE-0943441. S.J.G. acknowledges support from the DFG (GL 203/6-1), SFB 631 and the Fonds der
Chemischen Industrie. Experiments were performed at the Bavarian NMR center at TU
M\"unchen.






\begin{thebibliography}{00}


\bibitem{Hill73}
Barrett L. Tomlinson and H. D. W. Hill, Fourier synthesized excitation of nuclear magnetic resonance with application to homonuclear decoupling and solvent line suppression, J. Chem. Phys. 59,(1973) 1775--1785

\bibitem{Bod76}
Geoffrey Bodenhausen, Ray Freeman, Gareth A. Morris, A simple pulse sequence for selective excitation in Fourier transform NMR , J, Magn. Reson. 23 (1976) 171-175

\bibitem{Freeman78}
G. A. Morris and R. Freeman, Selective excitation in Fourier transform nuclear magnetic resonance, J. Magn. Reson. 29 (1978) 433--462.

\bibitem{Hoult79}
D.I. Hoult, The solution of the Bloch equations in the presence of a varying B1 field: An approach to selective pulse analysis, J. Magn. Reson. 35 (1979) 68--86.

\bibitem{Bod82}
Lyndon Emsley, Geoffrey Bodenhausen, Self-refocusing effect of 270° Gaussian pulses. Applications to selective two-dimensional exchange spectroscopy, J. Magn. Resona. 82 (1989) 211--221

\bibitem{Ernst83}
P. Caravatti, G. Bodenhausen, R. R. Ernst, Selective pulse experiments in high-resolution solid state NMR, J. Magn. Reson. 55 (1983) 88--103

\bibitem{Freeman84}
Christopher Bauer, Ray Freeman, Tom Frenkiel, James Keeler, A. J. Shaka, Gaussian pulses, J. Magn. Reson. 58 (1984) 442--457

\bibitem{Silver&Hoult84}
M.S. Silver, R.I. Joseph, D.I. Hoult, Highly selective $\pi$/2 and $\pi$ pulse generation. J. Magn. Reson. 59 (1984) 347--351.

\bibitem{Freeman87}
Jan Friedrich, Simon Davies, Ray Freeman, Shaped selective pulses for coherence-transfer experiments, J. Magn. Reson. 75 (1987) 390--395

\bibitem{Hardy88}
C.J. Hardy, P.A. Bottomley, M. O'Donnell, and P. Roemer, Optimization of two dimensional spatially selective NMR pulses by simulated annealing, J. Magn. Reson. 77 (1988) 223--250. 

\bibitem{Warren88}
F. Loaiza, M. A. McCoy, S. L. Hammes, W. S. Warren, Selective excitation without phase distortion using self-refocused amplitude- and amplitude/phase-modulated pulses, J. Magn. Reson. 77 (1988) 175--181

\bibitem{Canet88}
Piotr Tekely, Jean Brondeau, Karim Elbayed, Alain Retournard and, Daniel Canet, A simple pulse train, using 90° hard pulses, for selective excitation in high-resolution solid-state NMR, J. Magn. Reson. 80 (1988) 509--516

\bibitem{Freeman89a}
X.-L. Wu, R. Freeman, Darwin's ideas applied to magnetic resonance. The marriage broker, J. Magn. Reson. 85 (1989) 414--420

\bibitem{Freeman89b}
H. Geen, S. Wimperis, R. Freeman, Band-selective pulses without phase distortion. A simulated annealing approach, J. Magn. Reson. 85 (1989) 620--627

\bibitem{Bod89}
L. Emsley and G. Bodenhausen, Self-refocusing 270° gaussian pulses for slice selection without gradient reversal in magnetic resonance imaging, Magn, Reson. Med. 10 (1989) 273--281

\bibitem{Yagle90}
A.E. Yagle, Inversion of the Bloch transform in magnetic resonance imaging using asymmetric two-component inverse scattering, Inverse Problem 6 (1990) 133--151. 

\bibitem{Shinnar89a}
M. Shinnar, L. Bolinger, J. S. Leigh, The synthesis of pulse sequences yielding arbitrary magnetization vectors, Mag. Reson. Med. 12 (1989) 74--80.

\bibitem{Shinnar89b}
M. Shinnar, L. Bolinger, J. S. Leigh, The use of finite impulse response filters in pulse design, Mag. Reson. Med. 12 (1989) 81--87.

\bibitem{Shinnar89c}
M. Shinnar, L. Bolinger, J. S. Leigh, The synthesis of pulse sequences yielding arbitrary magnetization vectors, Mag. Reson. Med. 12 (1989) 78--92.

\bibitem{Shinnar89d}
M. Shinnar, J. S. Leigh, 
The application of spinors to pulse synthesis and analysis, Mag. Reson. Med. 12 (1989) 93--98.

\bibitem{Pauly91}
J.~Pauly, P.~Le Roux, D.~Nishimur, A.~Macovski, Parameter relations for the Shinnar-Le Roux selective excitation pulse design algorithm, IEEE Trans.~Med.~Imag.~ 10 (1991) 53--65.

\bibitem{Carlson91}
J.W.Carlson, Exact solution for selective-excitation pulses. J. Magn. Reson. 94 (1991) 376--386.

\bibitem{Freeman91a}
H. Geen and R. Freeman, Band-selective radiofrequency pulses . J. Magn. Reson. 93 (1991) 93-141.

\bibitem{Freeman91b}
R. Freeman, Selective excitation in high-resolution NMR, Chem. Rev. 91-7 (1991) 1397--1412

\bibitem{Freeman92}
R. Freeman, High resolution NMR using selective excitation, J. Mol. Struct, 266 (1992) 39--51

\bibitem{Carlson92}
J.W.Carlson, Exact solution for selective-excitation pulses. II Excitation pulses with phase control. J. Magn. Reson.97 (1992) 65--78.

\bibitem{Rourke92}
D.E.Rourke and P.G.Morris, The inverse scattering and its use in the exact inversion of the Bloch equation for noninteracting spins. J. Magn. Reson. 99 (1992) 118--138.

\bibitem{Bod92}
L. Emsley, G. Bodenhausen, Optimization of shaped selective pulses for NMR using a quaternion description of their overall propagators, J. Magn. Reson. 97 (1992) 135--148

\bibitem{K&F93}
\=E. Kup\v{c}e, R.Freeman, Polychromatic selective pulses, J. Magn. Reson. Ser.A 102 (1993) 122--126.

\bibitem{Rourke94}
D.E. Rourke, M.J.W. Prior, P.G. Morris, J.A.B. Lohman, Stereographic projection method of exactly calculating selective pulses, J. Magn. Reson. A 107 (1994) 203--214.

\bibitem{Emsley94}
L. Emsley, Selective pulses and their applications to assignment and structure determination in nuclear magnetic resonance, Methods Enzymol. 239 (1994) 207--246

\bibitem{K&F95}
\=E. Kup\v{c}e, R. Freeman, Band-Selective Correlation Spectroscopy, J. Magn. Reson. A 112 (1995)  134--137

\bibitem{Kupce95}
\=E. Kup\v{c}e, J. Boyd, I.D. Campbell, Short selective pulses for biochemical applications, \JMRB\ 106 (1995) 300--303.

\bibitem{Emsley95}
S. Caldarelli, A. Lesage, L. Emsley, Pure-phase selective excitation in NMR by acquisition during the pulse, J. Magn. Reson. A 116 (1995) 129--132

\bibitem{Boehlen96}
C. Dalvit, S. Y. Ko, J.-M. B\"ohlen, Single and multiple-selective excitation combined with pulsed field gradients, J. Magn. Reson. B 10 (1996) 124--131

\bibitem{Emsley96}
P. Borgnat, A. Lesage, S. Caldarelli, L. Emsley, Narrowband linear selective pulses for NMR, J. Magn. Reson. A 119 (1996) 289--294

\bibitem{Lunati98}
E. Lunati, P. Cofrancesco, M. Villa, P. Marzola, F. Osculati, Evolution Strategy Optimization for Selective Pulses in NMR, J. Magn. Reson. 134 (1998) 223--235

\bibitem{Virgili98}
T. Parella, F. Sánchez-Ferrando, A. Virgili, A simple approach for ultraclean multisite selective excitation using excitation sculpting, J. Magn. Reson. 135 (1998) 50--53

\bibitem{Freeman98}
Ray Freeman, Shaped radiofrequency pulses in high resolution NMR, Prog. Nucl. Magn. Reson. Spectrosc. 32, (1998) 59--106

\bibitem{Freeman99}
P. Xu, X.-L. Wu, R. Freeman, User-friendly selective pulses, J. Magn. Reson. 99 (1992) 308--322

\bibitem{Mueller99}
M. A. McCoy, L. Mueller, Nonresonant effects of frequency-selective pulses, J. Magn. Reson. 99 (1992) 18--36

\bibitem{Canet99}
C. Roumestand,C. Delay, J. A. Gavid, D. Canet, A practical approach to the implementation of selectivity in homonuclear multidimensional NMR with frequency selective-filtering techniques. Application to the chemical structure elucidation of complex oligosaccharides (pages 451–478),  Magn. Reson. Chem. 37 (1999) 451--478.

\bibitem{Canet00}
C. Roumestand, D. Canet, Extending the excitation sculpting concept for selective excitation, J. Magn. Reson. 147 (2000) 331--339

\bibitem{Sterk01}
K. Zangger, M. Oberer, H. Sterk, Pure-phase selective excitation in fast-relaxing systems, J. Magn. Reson. 152 (2001) 48--56

\bibitem{Emsley02}
P. Charmont, D. Sakellariou, L. Emsley, Sample restriction using radiofrequency field selective pulses in high-resolution solid-state NMR, J. Magn. Reson. 154 (2002) 136--141

\bibitem{Shaka02}
K. E. Cano, M. A. Smith, A. J. Shaka, Adjustable, broadband, selective excitation with uniform phase, J. Magn. Reson. 155 (2002) 131--139

\bibitem{Griffin04}
M.~Veshtort, R.~Griffin, High-performance selective excitation pulses for solid-and liquid-state NMR spectroscopy, Chem.~Phys.~Chem.~ 5 (2004) 834--850.


\bibitem{Hajduk93}
P. J. Hajduk, D. A. Horita, and L. E. Lerner, Theoretical analysis of relaxation during shaped pulses. I The effects of short T$_1$ and T$_2$, \JMRA \vol{103} (1993) 40--52.

\bibitem{Raddi2000}
A. Raddi and U. Klose, Relaxation effects on transverse magnetization using RF pulses long compared to T2, \JMR\ 144 (2000) 108--114.

\bibitem{Nuzillard94}
J.-M. Nuzillard and R. Freeman, Band-selective pulse designed to accommodate relaxation, \JMRA\ 107 (1994) 113--118.

\bibitem{Shen95}
J.~Shen,L.~E.~Lerner, J.~Magn.~Reson. A, Selective radiofrequency pulses minimizing relaxation, 112 (1995) 265--269.

\bibitem{Rourke04}
D.E. Rourke, L. Khodarinova, and A.A. Karabanov, Two-Level Systems with Relaxation, Phys. Rev. Letters, 92 (2004), 163003(1--4). 

\bibitem{Rourke07}
D.E. Rourke, A.A. Karabanov, G.H. Booth, I. Frantsuzov, The Bloch equation when T1 = T2, Inverse Problems 23 (2007) 609--623.

\bibitem{Issa09}
B. Issa, Design of self-refocused pulses under short relaxation times, \JMR\ 198 (2009) 151--159.

\bibitem{Khaneja01}
N. Khaneja, R. Brockett, S. J. Glaser, Time optimal control in spin systems, 
Phys. Rev. A 63 (2001) 032308-13. 

\bibitem{Reiss02}
T. O. Reiss, N. Khaneja, S. J. Glaser,
Time-optimal coherence-order-selective transfer 
of in-phase coherence in heteronuclear IS 
spin systems, \JMR\ 154 (2002) 192--195.

\bibitem{Skinner03}
T.E. Skinner, T.O. Reiss, B. Luy, N. Khaneja, S.J. Glaser, Application of optimal control theory to the design of broadband excitation pulses for high resolution NMR, \JMR \vol{163} (2003) 8--15.

\bibitem{Khaneja03}
N. Khaneja,  T. Reiss, B. Luy, S. J. Glaser,
Optimal control of spin dynamics in the presence of relaxation,
\JMR\ 162 (2003) 311--319.

\bibitem{Skinner04}
T.E. Skinner, T.O. Reiss, B. Luy, N. Khaneja, S.J. Glaser, Reducing the duration of broadband excitation pulses using optimal control with limited RF amplitude, \JMR \vol{167} (2004) 68--74.

\bibitem{Kobzar04}
K. Kobzar, T.E. Skinner, N. Khaneja, S.J. Glaser, B. Luy, Exploring the limits of broadband excitation and inversion pulses, \JMR\ 170 (2004) 236--243.
17,

\bibitem{Skinner05}
T.E. Skinner, T.O. Reiss, B. Luy, N. Khaneja, S.J. Glaser, Tailoring the optimal control cost function to a desired output: application to minimizing phase errors in short broadband excitation pulses, \JMR\ 172 (2005) 17--23. 

\bibitem{Grape}
N. Khaneja, T. Reiss, C. Kehlet, T. Schulte-Herbr\"{u}ggen, and S.J. Glaser, Optimal control of coupled spin dynamics: design of NMR pulse sequences by gradient ascent algorithms, \JMR \vol{172} (2005) 296--305.

\bibitem{Kobzar05}
K. Kobzar, B. Luy, N. Khaneja, S. J. Glaser,
Pattern pulses: Ddsign of arbitrary excitation profiles as a function of pulse amplitude and offset,
\JMR\ 173 (2005) 229--235.

\bibitem{Skinner06}
T.E. Skinner, K. Kobzar, B. Luy, M.R. Bendall, N. Khaneja, S.J. Glaser Optimal control design of constant amplitude phase-modulated pulses: Application to calibration-free broadband excitation, \JMR\ 179 (2006) 241--249.

\bibitem{Neves06}
J. L. Neves, B. Heitmann, T. O. Reiss, H. H. R. Schor, N. Khaneja, S. J. Glaser,
Exploring the limits of polarization transfer efficiency in homonuclear three spin systems,
\JMR\ 181 (2006) 126--134.

\bibitem{Gershenzon07}
N.I. Gershenzon, K. Kobzar, B. Luy, S.J. Glaser, T.E. Skinner, Optimal control design of excitation pulses that accommodate relaxation, \JMR\ 188 (2007) 330--336.

\bibitem{Gershenzon08}
N. I. Gershenzon, T. E. Skinner, B. Brutscher, N. Khaneja, M. Nimbalkar, B. Luy, and S. J. Glaser,    Linear phase slope in pulse design: Application to coherence transfer,” J. Magn. Reson. 192 (2008) 235--243. 

\bibitem{Kobzar08}
K. Kobzar, T. E. Skinner, N. Khaneja, S. J. Glaser, B. Luy, Exploring the limits of broadband excitation and inversion: II. RF-power optimized pulses,"  J. Magn. Reson. 194 (2008) 58--66 .

\bibitem{Neves09}
J. L. Neves, B. Heitmann, N. Khaneja, S. J. Glaser,
Heteronuclear decoupling by optimal tracking, \JMR\ 201 (2009) 7--17.

\bibitem{Braun10}
M. Braun, S. J. Glaser,
Cooperative pulses,
\JMR\ 207 (2010) 114--123.

\bibitem{Glaser10}
N. Pomplun,  S. J. Glaser,
Exploring the limits of electron-nuclear polarization transfer efficiency in three spin systems,
Phys.~Chem.~Chem.~Phys.~ 12 (2010) 5791--5798.

\bibitem{Skinner11}
T. E. Skinner, M. Braun, K. Woelk, N. I. Gershenzon, and S. J. Glaser, Design and application of robust RF pulses for toroid cavity NMR spectroscopy,”  J. Magn. Reson. 209 (2011) 282--290.

\bibitem{Glaser11}
Y. Zhang, M. Lapert, D. Sugny, M. Braun, S. J. Glaser,
Time-Ootimal control of spin 1/2 particles in the presence of
radiation damping and relaxation,
J.~Chem.~Phys. 134 (2011) 054103.

\bibitem{SkinnerENC07}
T. E. Skinner, N. I. Gershenzon, B. Luy, and S. J. Glaser, Optimal control of band-selective excitation pulses to accommodate relaxation,” 48th Experimental NMR Conference (ENC), April, 2007, Daytona Beach, Florida.

\bibitem{GershenzonENC07}
N. I. Gershenzon, S. J. Glaser, and T. E. Skinner, Empirical relations for the design of band-selective pulses using optimal control theory,” 48th Experimental NMR Conference (ENC), April, 2007, Daytona Beach, Florida.

\bibitem{Lurie85}
D. J. Lurie, A systematic design procedure for selective pulses in NMR imaging, Magn. Reson. Imaging 3 (1985) 235--243. 

\bibitem{O'Donnell85}
M. O’Donnell and W. J. Adams, Selective time-reversal pulses for NMR imaging, Magn. Reson. Imaging 3 (1985) 377--382.

\bibitem{Conolly86}
S. Conolly, D. Nishimura, A. Macovski, Optimal control solutions to the magnetic resonance selective excitation problem, IEEE Trans. Med. Imaging MI-5 (1986) 106--115.

\bibitem{Mao86}
J. Mao, T.H. Mareci, K.N. Scott, E.R. Andrew, Selective inversion radiofrequency pulses by optimal control, \JMR\ 70 (1986) 310--318.

\bibitem{Murdoch87}
J. B. Murdoch, A. H. Lent, and M. R. Kritzer, Computer-optimized narrowband pulses for multi-slice imaging, \JMR\ 74 (1987) 226--263.

\bibitem{Rosenfeld96}
D. Rosenfeld, Y. Zur, Design of adiabatic selective pulses using optimal control theory, \MRM\ 36 (1996) 401--409. 

\bibitem{Rosenfeld97}
D. Rosenfeld, S. L. Panfil, Y. Zur, Optimization of adiabatic selective pulses, J. Magn. Reson. 126 (1997) 221--228

\bibitem{James91a} 
V. Smith, J. Kurhanewicz, and T. L. James, Solvent-suppression pulses. I. Design using optimal control theory, \JMR\ 95 (1991) 41--60.

\bibitem{Schulte07}
R.F.Schulte, A. Hennig, J. Tsao, P. Boesiger, K.P.Pruessmann, Design of broadband RF pulses with polynomial-phase response, \JMR\ 186 (2007) 167--175.

\bibitem{Schulte08}
R.F.Schulte, P. Le Roux, M. W. Vogel, H. Koenig, Design of phase-modulated broadband refocusing pulses, \JMR\ 190 (2008) 271--279.

\bibitem{Rabiner}
L.R.Rabiner, B. Gold, Theory and Application of Digital Signal Processing, Prentice-Hall, Englewood Cliffs, N.J.(1975).

\bibitem{Parks}
T.W.Parks, C.S. Burrus, Digital Filter Design, John Wiley and Sons, New York (1987).



%
%

%

\end{thebibliography}



\vfill\eject

\par\vfill\eject

\par\vfill\eject

\par\vfill\eject

\par\vfill\eject

\end{document}